\documentclass[10.5pt, a4paper]{IEEEtran}
\IEEEoverridecommandlockouts

\usepackage{xspace}
\usepackage{lipsum}
\usepackage{algpseudocode}
\usepackage{algorithm}
\usepackage{mathtools}
\usepackage{amssymb}
\usepackage{amsmath}
\usepackage{multirow}
\usepackage{pifont}
\usepackage{cuted}
\usepackage{tabularray}
\usepackage{float}

\usepackage{stfloats}

\usepackage{comment}

\usepackage{cite}

\usepackage{soul}
\usepackage{color}
\usepackage{graphics}
\usepackage{subfig}
\usepackage{tablefootnote}
\usepackage{threeparttable}
\usepackage[symbol]{footmisc}
\usepackage{caption}
\usepackage{siunitx}

\newcolumntype{P}[1]{>{\centering\arraybackslash}p{#1}}

\usepackage{hyperref}
\hypersetup{
    pdfcreator={},   
    pdfproducer={},  
}

\usepackage{mathtools}

\def\BibTeX{{\rm B\kern-.05em{\sc i\kern-.025em b}\kern-.08em
    T\kern-.1667em\lower.7ex\hbox{E}\kern-.125emX}}

\newcommand{\ATNTT}{\texttt{@NTT}\xspace}    
\begin{document}

\title{
\ATNTT: \underline{A}lgorithm-\underline{T}argeted NTT hardware acceleration via Design-Time Constant Optimization


}

\author{\IEEEauthorblockN{Mohammed Nabeel, Mahmoud Hafez, Michail Maniatakos}
\IEEEauthorblockA{
\\
\textit{New York University Abu Dhabi (NYUAD), Abu Dhabi, UAE}
}
}

\maketitle

\begin{abstract}
The Number Theoretic Transform (NTT) is a critical computational bottleneck in many lattice-based post-quantum cryptographic (PQC) algorithms. By leveraging the Fast Fourier Transform (FFT) algorithm, the NTT of a polynomial of degree~$N-1$ can be computed with a time complexity of~$\mathcal{O}(N \log N)$. Hardware implementation of NTT is generally preferred over software ones, as the latter are significantly slower due to complex memory access patterns and modular arithmetic operations. Achieving maximum throughput in hardware, however, typically demands a prohibitively large number of butterfly unit instantiations. In this work, we propose \ATNTT, which exploits the fact that the ring parameters in these algorithms are fixed, enabling design-time constant optimization and achieving the maximum throughput of $N$-point NTT per clock cycle with a compact hardware footprint.
Our case study on the Dilithium NTT, implemented using the TSMC 28\,nm library, operates at a clock frequency of $1.0~\mathrm{GHz}$ with an area of $1.45~\mathrm{mm^2}$. On FPGA, the design achieves a throughput-per-LUT that is $5.2\times$ higher than the state-of-the-art implementation.

\end{abstract}
\section{Introduction}
As advancements in quantum computers gain momentum, widely used public-key cryptography schemes such as RSA and ECC are at risk \cite{shor1999polynomial}. To protect the public key infrastructure from such quantum-computer attacks, the National Institute of Standards and Technology (NIST) has started standardizing replacements for RSA and ECC. These replacement algorithms are commonly referred to as Post-Quantum Cryptography (PQC) algorithms.

Most of the currently standardized PQC algorithms selected by NIST for key exchange and digital signatures are lattice-based, relying on the Ring-LWE or Module-LWE problem. In these schemes, elements are represented as polynomials of degree \(N-1\) over the ring $R_Q =  \mathbb{Z}_Q[X]/(X^N+1)$, where \(Q\) is the coefficient modulus. The most complex arithmetic operation in $R_Q$ is polynomial multiplication. Since polynomial multiplication in the point-value representation can be performed using simple Hadamard multiplication, it is common to transform the coefficient representation of a polynomial to its point-value representation using NTT.

Table~\ref{tab:PQCalgo} lists the standardized algorithms from NIST that require NTT during computation. The NTT contributes to the majority of the execution time in Kyber and Dilithium. Falcon employs both Discrete Fourier Transform (DFT) and NTT implementations, and together they account for a significant portion of the execution time dominated by transform operations. Accelerating NTT is critical for these algorithms' performance. Hardware acceleration is preferred over software due to the modular arithmetic involved and the complex memory access patterns during NTT, which make software implementations significantly slower.

\begin{table}[!t]
\centering
\caption{Standardized PQC algorithms that use NTT and their Parameters (NIST 2025).}
\begin{tabular}{l|l|c}
\hline
\textbf{Algorithm} & \textbf{Usage} & \textbf{Ring Parameters (Q, N)} \\
\hline \hline
ML-KEM (Kyber)     & KEM/Encryption & (3329, 256)       \\
ML-DSA (Dilithium) & Signature      & (8380417, 256)        \\
FN-DSA (FALCON)    & Signature      & (12289, 512/1024)   \\
\hline
\end{tabular}
\label{tab:PQCalgo}
\end{table}

To this end, researchers have developed hardware accelerators for these PQC algorithms, consistently identifying the NTT as their most computationally expensive operation \cite{YamanKyberDATE, NiNoBRAMKyberISCAS, zhaoDilithiumTCCHES, guptaDilithiiumTCAS1, DamFalconISCAS, yaman2021hardware}. For example, \cite{zhaoDilithiumTCCHES} reports that a 256-point NTT in Dilithium requires 552 clock cycles, while \cite{zhaoDilithiumTCCHES} achieves the same operation in 606 clock cycles.

In this work, we present \ATNTT, a framework designed to achieve high throughput of N-point NTT computations per clock cycle while maintaining lower power consumption and higher efficiency per unit area for the overall computation.
To achieve this, we exploit the fact that the ring parameters of these algorithms are fixed in most cases, except for some schemes like Falcon, which use different values of $N$ for 128-bit and 256-bit security.
Instead of storing them as constants in registers and memories, we define them as synthesis-time constants ($pararmeter$ in Verilog, $generic$ in VHDL). This not only enables packing more computation units into a smaller area but also reduces storage requirements and memory access complexity that would otherwise be incurred. The main contributions of the work are as follows:
\begin{itemize}
    \item We propose \ATNTT, a framework that pushes the limits of NTT throughput to achieve N-point NTT computation per clock cycle while maintaining low power consumption and high area efficiency.
    \item We evaluate \ATNTT both for  ASIC and FPGA targets.
    \item We optimally decompose each constant multiplier required for the NTT into a minimal set of shifts and adders/subtractors, achieving significantly better optimization than industry-standard synthesis tools.
    \item We will open-source the artifacts.
\end{itemize}

\section{Background}
\subsection{NTT}
\label{sec:NTT}
\begin{figure}[!t]
    \centering
    \includegraphics[width=.85\linewidth]{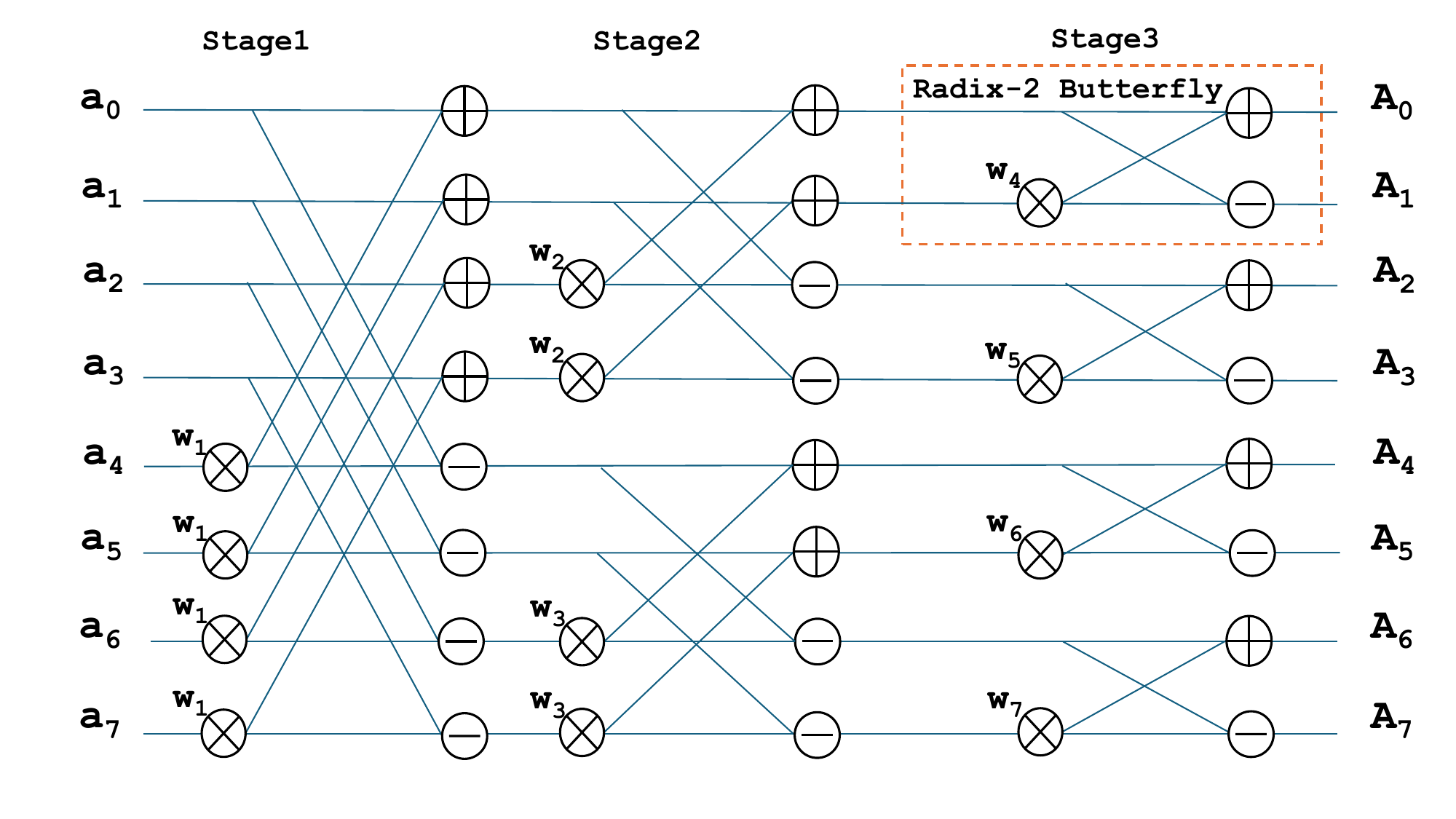}\vspace{-0.1cm}
    \caption{\textbf{NTT Flow.} 8-point NTT flow. Each of the $log_28$ stages involves $\frac{8}{2}$ radix-2 butterfly operations. For the fastest execution, one needs to implement all these radix-2 butterflies.}
    \label{fig:NTTflow}
\end{figure}

The Number Theoretic Transform (NTT) ~\cite{NTTscott2017note} is a generalized form of the Discrete Fourier Transform (DFT). NTT can be seen as the DFT on a ring over a finite field, where the ring is an integer modulo a prime number $Q$. As with the complex root of unity in DFT, for NTT, it is the primitive $N^{th}$ root of unity $w$, such that $w^N = 1 \mod Q$, where $N$ is the polynomial degree. The primitive $N^{th}$ root of unity is also known as the \textit{twiddle factor}. The NTT enables efficient conversion between the coefficient and point-value representations of polynomials using the FFT algorithm, reducing the complexity of polynomial multiplication from $\mathcal{O}(N^2)$ to $\mathcal{O}(N \log N)$. An 8-point NTT is depicted in Fig. \ref{fig:NTTflow}. The basic computation unit for NTT is a radix-2 NTT operation, also called the butterfly operation.

For an N-point NTT operation there will be a total of $\frac{N}{2} log_2~N$ butterfly operations. Inverse NTT (iNTT) operation, to convert the point-representation back to the coefficient also can be implemented using forward-NTT flow shown in Fig. \ref{fig:NTTflow}, but the input elements has to be loaded in bit-reversed\footnote{Bit-reversed order refers to reindexing an array such that each index is replaced by the value obtained by reversing the binary representation of that index.} order and the twiddle factor for a given butterfly will be the modular inverse ($w^{-1} \pmod{Q}$) of the twiddle that is loaded for forward-NTT. Also, in iNTT, either all input elements or all output elements must be multiplied by $N^{-1} \pmod{Q}$.

\subsection{Butterfly operation}
Figure~\ref{fig:Rad2Bfly} illustrates a typical implementation of the butterfly unit. The butterfly unit performs the radix-2 NTT operation. The bulk of the logic is dedicated to the modular multiplication of $B \cdot W \bmod Q$. Efficient reduction modulo $Q$ after the multiplication of $B \cdot W$ is critical for performance. Two common methods are \emph{Barrett} and \emph{Montgomery reduction}~\cite{cryptohandbook}, both avoiding costly division but with different trade-offs. For our analysis, we assume \emph{Barrett reduction} for modular reduction after multiplication, since Montgomery reduction requires expensive domain conversion of at least one operand. As shown in Fig.~\ref{fig:Rad2Bfly}, following the regular multiplication (Mult1), \emph{Barrett reduction} requires two additional multiplication operations (Mult2 and Mult3) to perform the modular reduction. Among the various arithmetic circuits, these multipliers constitute the primary contributors to area, power consumption, and critical path delay.

%
\section{\ATNTT Framework}

\begin{figure}[t]
    \centering
    \includegraphics[width=.85\linewidth]{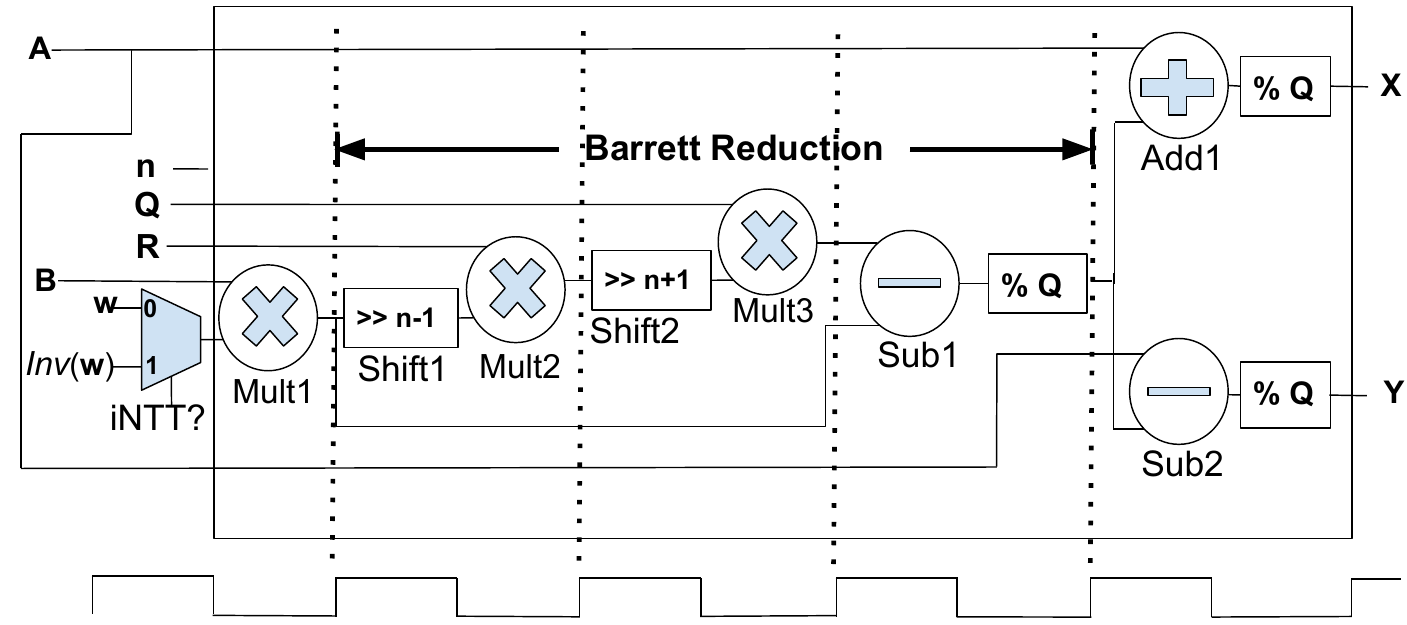}\vspace{-0.1cm}
    \caption{\textbf{Radix-2 Butterfly.} Computes $A + B\cdot $w$ \pmod Q$ and $A - B \cdot $w$ \pmod Q$. $A,B \in \mathbb{Z}_Q$; $n = \lceil \log_2 Q \rceil$; $R = \lfloor 4^n / Q \rfloor$  }
    \label{fig:Rad2Bfly}
\end{figure}

This section details the process undertaken to optimize the basic processing element (PE) used for NTT computation, the radix-2 butterfly circuit shown in Fig.\ref{fig:Rad2Bfly}. The focus is on leveraging operands, which are fixed for the targeted PQC algorithm as determined by the ring parameters (Table~\ref{tab:PQCalgo}), to enhance hardware accelerators' efficiency, with the goal of fitting more PEs per unit area and thereby improving the throughput of NTT.

\begin{table}[t]
\centering
\caption{Fixed values in PQC algorithms and their corresponding compute units optimized.}
\label{tab:fixed_values}
\begin{tabular}{c|c|c}
\hline
\textbf{Fixed Value} & \textbf{Description} & \textbf{Compute Unit Optimized} \\ \hline \hline
$Q$      & Coefficient modulus          & Mult3, \%~Q \\ 
$n$      & $\log_2 Q$                   & Shift1, Shift2 \\ 
$R$      & $\lfloor 4^n / Q \rfloor$    & Mult2 \\ 
$w$      & $w^N= 1\pmod{Q}$             & Mult1 \\
$N^{-1}$ & Normalization factor (iNTT)  & Multiplier per element \\ \hline
\end{tabular}
\end{table}

In Fig.~\ref{fig:Rad2Bfly}, except for the operands \(A\) and \(B\), all other inputs can be precomputed since they depend solely on the ring parameters, the polynomial degree \((N - 1)\) and the coefficient modulus \(Q\). Traditionally, these constant values have been stored in registers or memory. However, treating these precomputed values as synthesis-time constants allows for further optimization of the arithmetic units, particularly the multipliers (Mult1, Mult2 and Mult3), which are the dominant contributors to area, power consumption, and critical path delay.

These multipliers become constant multipliers. One option is to rely on the synthesis tool for constant-multiplication optimization. However, our analysis shows that, this introduces significantly higher overhead (10\% for ASIC and 20\% for FPGA) compared to explicitly decomposing constant multipliers into the minimal number of shifts and add/sub operations. The multiplication of a constant by a variable is generally implemented using a shift-and-add architecture, employing only shift and addition/subtraction operations~\cite{nguyen2000number} (Ex: $13 \cdot w = w << 3 + w << 2 + w$, 2 adders and 2 shifts instead of regular multiplication). Note that shifts can be realized using only wires, which incur no hardware cost.

\begin{figure}[t]
    \centering
    \includegraphics[scale=.9]{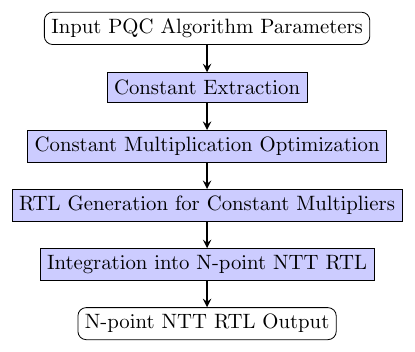}\vspace{-0.1cm}
    \caption{{\ATNTT flow.}}
    \label{fig:Flow_diag}
\end{figure}

Thus, the optimization problem is to find the minimum number of adders/subtractors required to implement the constant multiplication.
Although this optimization problem has been investigated in prior work~\cite{voronenko2007multiplierless,spiralMultless,de2019table,lagoon2020deriving}, \ATNTT builds upon these foundations and incorporates key ideas from~\cite{voronenko2007multiplierless} within a new end-to-end framework tailored for the NTT setting used in the target PQC algorithm. The core of the framework is to identify all constants used in the multipliers of the target PQC algorithm, optimize these constant multiplications, generate the corresponding RTL code, and integrate them to form an N-point NTT RTL design. The high-level steps involved in \ATNTT are summarized in Fig.~\ref{fig:Flow_diag}.
The high-level steps involved in \ATNTT are summarized in Fig.~\ref{fig:Flow_diag}.

As discussed in Section~\ref{sec:NTT}, to support the iNTT using the same butterfly hardware, the modular inverse of the twiddle factor must be supplied as the input when operating in iNTT mode, as shown in Fig.~\ref{fig:Rad2Bfly}. In other words, Mult1 performs a multiple-constant multiplication (MCM). 
Finding an optimal MCM solution, i.e., one that minimizes the number of additions and subtractions, is known to be NP-complete. Motivated by this, \ATNTT adopts the near-optimal strategy of~\cite{voronenko2007multiplierless} as a core component within our overall optimization flow.
Another constant multiplication required for iNTT is the multiplication of all the input or output elements with $N^{-1} \pmod{Q}$.

In addition to the improvements in power, performance, and area, constant multiplication also reduces the critical-path delay, which in turn allows for fewer pipeline stages, further lowering area and power and reducing the initiation interval (II). Typically, pipeline registers are inserted after every multiplication (Fig.~\ref{fig:Rad2Bfly}), since the multiplier constitutes the critical path.

Moreover, twiddle factors are typically generated on the fly using dedicated hardware or stored in internal memories or register files. With constant multiplication optimization applied, however, the twiddle factors are directly merged into the design logic, eliminating the need for their generation or storage and the associated data movement logic, thereby contributing to area and power savings.

Although this work performs the analysis using a modular multiplier based on Barrett reduction to identify constant multipliers and convert them to an optimized form, the \ATNTT framework can be easily adapted for other modular multipliers.

\section{Evaluation}

\begin{table}[t]
\centering
\caption{Radix-2 butterfly without any constant optimization Vs.\\ when implemented using \ATNTT}
\label{tab:butterfly_comparison}
\begin{tabular}{l|r|r|c|l}
\hline
                     & \multicolumn{2}{c|}{\textbf{ASIC}}             & \multicolumn{2}{c}{\textbf{FPGA}}  \\ \cline{2-5} 
                     & \textbf{Area}              & \textbf{Freq}     & \textbf{LUT/REG/} & \textbf{Freq}               \\
 \textbf{NTT-Design} & \textbf{$\mu m^2$}       & \textbf{$Ghz$}  & \textbf{DSP}      & \textbf{$Mhz$} \\ \hline \hline
Kyber (no opt)       & 1872                       & 1.0               & 253~/~94~/~3      & 356 \\ 
Kyber (\ATNTT)       & 754                        & 1.0               & 159~/~109~/~0     & 451 \\ \hline \hline
Dilithium (no opt)   & 4995                       & 1.0               & 480~/~229~/~5     & 229 \\ 
Dilithium (\ATNTT)   & 1532                       & 1.0               & 305~/~182~/~0     & 305 \\ \hline
\end{tabular}
\end{table}

\begin{table}[t]
\centering
\caption{256-point NTT with synthesis-time constant optimization Vs.\\ when implemented using \ATNTT}
\label{tab:asic_fpga_comparison}
\begin{tabular}{l|r|r|c|l}
\hline
                      & \multicolumn{2}{c|}{\textbf{ASIC}}                   & \multicolumn{2}{c}{\textbf{FPGA}}          \\ \cline{2-5} 
                      & \textbf{Area}               & \textbf{Freq}          & \textbf{LUT/REG/}       & \textbf{Freq}    \\
 \textbf{NTT-Design}  & \textbf{$mm^2$}        & \textbf{$Ghz$}       & \textbf{DSP}            & \textbf{$Mhz$} \\ \hline \hline
Kyber (Synth-opt)     & 0.70                         & 1.0                & 198295~/~97816~/~0      & 433 \\ 
Kyber (\ATNTT)        & 0.62                         & 1.0                & 142719~/~97815~/~0      & 451 \\ \hline \hline
Dilithium(Synth-opt)  & 1.53                         & 1.0                & 385693~/~187567~/~0     & 325 \\ 
Dilithium (\ATNTT)    & 1.45                         & 1.0                & 311500~/~187178~/~0     & 305 \\ \hline
\end{tabular}
\end{table}


\begin{table*}[t]
\centering
\caption{NTT throughput Comparison for Dilithium and Kyber with the SoTA}
\label{tab:ntt_comp}
\begin{tabular}{l l c r r r r r r}
\hline
\textbf{Scheme} & \textbf{Work} & \textbf{FPGA}  & \textbf{Freq ($Mhz$)} & \textbf{CC} & \textbf{LUTs} & \textbf{Latency ($\mu s$)} & \textbf{NTT/ms} & \textbf{(NTT/$ms$)/LUT}
\\ \hline\hline

Dilithium & \cite{nguyen2024high} & Artix-7 & 180 & 128 & 7451     & 0.7111 & 1406    & 0.19 \\
          & \ATNTT                & XCU50  & 305 & 1   & 311500   & 0.0033 & 305000  & 0.98 \\ \hline \hline

Kyber     & \cite{nguyen2000number} & Artix-7 & 250 & 277 & 2466   & 1.108 & 903    & 0.37 \\
          & \ATNTT                  & XCU50   & 451 & 1   & 142719 & 0.002 & 451000 & 3.16 \\ \hline

\end{tabular}
\end{table*}

In this section, we evaluate \ATNTT for Kyber and Dilithium NTTs. The NTT parameters for Kyber and Dilithium are given in Table~\ref{tab:PQCalgo}. Both require a 256-point NTT; however, Kyber performs the FFT operation over only seven stages instead of eight, and the twiddle factors used for Kyber are odd powers of the 
Nth root of unity rather than all powers of the  Nth root of unity \cite{Kyber_ntt}. Evaluation is done for both ASIC and FPGA targets. For ASIC, we use TSMC 28nm, and FPGA evaluation is performed on AMD Xilinx's XCU50 UltraScale FPGA. For ASIC synthesis, we use Synopsys DC and AMD Xilinx's Vivado for FPGA synthesis. Target frequency is set to $1~Ghz$ for ASIC and $500~Mhz$ for FPGA. All the designs are verified using Synopsys VCS functional simulation.
The baseline for the butterfly RTL is taken from the open-source, silicon-proven design presented in~\cite{CoFHEE, momalab2025cofhee}.

Table~\ref{tab:butterfly_comparison} presents the results of a non constant optimized butterfly unit and compares them with the average area of butterfly units after synthesizing the N point NTT design generated by the \ATNTT flow. Although this comparison is somewhat unfair since the non optimized design includes additional logic external to the butterfly such as data movement logic and storage or generation of twiddle factors while the \ATNTT butterfly has effectively absorbed and merged this functionality into the RTL, we still present this comparison to demonstrate that constant optimization alone already achieves significant saving even before accounting for any additional reductions at the system level.

On ASIC, \ATNTT reduces the area significantly, from $1872~\mu m^2$ to $754~\mu m^2$ for Kyber, and from $4995~\mu m^2$ to $1532~\mu m^2$ for Dilithium, demonstrating the effectiveness of constant multiplication optimization.

On FPGA, \ATNTT also reduces resource usage, especially LUTs, and eliminates the need for DSP blocks by absorbing constant multiplications into logic. For example, Kyber's LUT usage decreases from 253 to 159, while DSP usage drops from 3 to 0. Similarly, Dilithium's LUT usage drops from 480 to 305, with DSP blocks eliminated. The achieved operating frequency also remains high, resulting in an overall improvement in performance per area efficiency. These results illustrate that \ATNTT can generate highly area efficient constant-optimized butterfly units for NTT designs.

The results in Table~\ref{tab:asic_fpga_comparison} compare a 256-point NTT generated using \ATNTT against a baseline design relying only on synthesis-time constant optimization.
For Kyber, \ATNTT reduces LUT count by approximately 28\% while improving operating frequency by 4.2\%. On ASIC, the \ATNTT design achieves an 11–13\% area reduction while still meeting the target frequency of 1~$Ghz$. For Dilithium, \ATNTT yields a LUT reduction of 19\% on FPGA at the cost of a slight frequency drop, while ASIC area remains comparable to the synthesis-optimized version.

Importantly, the resulting designs fits comfortably within high-end FPGA devices such as the XCU50 (17\% utilization for Kyber and 36\% for Dilithium), highlighting the practicality of the generated RTL for deployment. Overall, these results demonstrate that \ATNTT provides tangible logic and area savings even when synthesis-time optimizations are already enabled, validating the benefit of performing constant optimization at RTL rather than relying on synthesis tools.

In Table~\ref{tab:ntt_comp}, we compare the NTT throughput against the state-of-the-art (SoTA) FPGA implementations for Kyber and Dilithium reported in~\cite{nguyen2024high}. Despite the increase in LUT usage due to full pipelining and constant-optimized datapaths, \ATNTT achieves a higher throughput-per-LUT efficiency, with improvements of $5.2\times$ for Dilithium and $8.5\times$ for Kyber.

\ATNTT achieves substantially higher throughput, producing up to $305{,}000$ NTT/$\mathrm{ms}$ for Dilithium and $451{,}000$ NTT/$\mathrm{ms}$ for Kyber on FPGA. In ASIC, the design delivers one N-point NTT every nanosecond. This is enabled by a deeply pipelined architecture that implements every NTT stage in hardware, increasing area but maximizing throughput.

\section{Conclusion}
In this work, we presented \ATNTT, an end-to-end framework for optimizing Number Theoretic Transform (NTT) architectures targeted to specific post-quantum cryptography (PQC) algorithms. By applying constant multiplication optimization and fully pipelined architecture generation, \ATNTT produces high-throughput, resource-efficient N-point RTL tailored to the target PQC algorithm. Compared to prior state-of-the-art designs, \ATNTT achieves substantial improvements in throughput while maintaining efficient area usage.

\bibliographystyle{IEEEtran}
\bibliography{references}

\end{document}